%% file: main.tex
\documentclass[sigconf, anonymous=False]{acmart}
\usepackage{booktabs} % For formal tables
\usepackage{multirow}
\usepackage{subcaption}
\usepackage{caption}
\usepackage{setspace}
\usepackage{amsmath}
\usepackage[english]{babel}
\definecolor{mygray}{gray}{0.5}
\usepackage{float}
\usepackage{mathtools}
\begin{document}

\copyrightyear{2017} 
\acmYear{2017} 
\setcopyright{acmcopyright}
\acmConference{SIGIR '17}{}{August 07--11, 2017, Shinjuku, Tokyo, Japan}\acmPrice{15.00}\acmDOI{http://dx.doi.org/10.1145/3077136.3080768}
\acmISBN{978-1-4503-5022-8/17/08}

\title{Word-Entity Duet Representations for Document Ranking}

\author{Chenyan Xiong}
\affiliation{
\institution{Language Technologies Institute\\ Carnegie Mellon University}
  \city{Pittsburgh} 
  \state{PA} 
  \postcode{15213}
  \country{USA}
}
\email{cx@cs.cmu.edu}

\author{Jamie Callan}
\affiliation{
\institution{Language Technologies Institute\\ Carnegie Mellon University}
  \city{Pittsburgh} 
  \state{PA} 
  \postcode{15213}
  \country{USA}
}
\email{callan@cs.cmu.edu}

\author{Tie-Yan Liu}
\affiliation{
\institution{Microsoft Research}
  \city{Beijing}  
  \postcode{100080}
  \country{P.R. China} 
}
\email{tie-yan.liu@microsoft.com}

\begin{abstract}
This paper presents a word-entity duet framework for utilizing knowledge bases in ad-hoc retrieval. 
In this work, the query and documents are modeled by word-based representations and entity-based representations.
Ranking features are generated by the interactions between the two representations, incorporating information from the word space, the entity space, and the cross-space connections through the knowledge graph.  To handle the uncertainties from the automatically constructed entity representations, an \texttt{att}ention-based \texttt{r}anking model \texttt{AttR-Duet} is developed.
With back-propagation from ranking labels, the model learns simultaneously how to demote noisy entities and how to rank documents with the word-entity duet. Evaluation results on TREC Web Track ad-hoc task demonstrate that all of the four-way interactions in the duet are useful, the attention mechanism successfully steers the model away from noisy entities, and together they significantly outperform both word-based and entity-based learning to rank systems.
\end{abstract}


\maketitle

\input{Introduction.tex}

\input{Related_work.tex}

\input{Duet.tex}

\input{Att_ltr.tex}

\input{Experiment.tex}

\input{Evaluation.tex}

\input{Conclusion.tex}

% \section{Acknowledgments}
% \clearpage
% \newpage
% \bibliographystyle{abbrv}
\bibliographystyle{ACM-Reference-Format}
\normalsize
% \fontsize{8.5pt}{8.5pt}\selectfont
% \vspace{0.1in}
\bibliography{citation}
\end{document}

%% file: Introduction.tex
\section{introduction}

Utilizing knowledge bases in text-centric search is a recent breakthrough in information retrieval~\cite{KG4IRTutorial}. The rapid growth of information extraction techniques and community efforts have generated large scale general domain knowledge bases, such as  DBpedia and Freebase. These knowledge bases store rich semantics in semi-structured formats and have great potential in improving text understanding and search accuracy.

There are many possible ways to utilize knowledge bases' semantics in different components of a search system. Query representation can be improved by introducing related entities and their texts to expand the query~\cite{daltonentity, xiong2015fbexpansion}. Document representation can be enriched by adding the annotated entities into the document's vector space model~\cite{raviv2016document,Xiong2016BOE,ESR}. The ranking model can also be improved by utilizing the entities and their attributes to build additional connections between query and documents~\cite{liu2015latent, EsdRank}. 
The rich choices of available information and techniques raise a new challenge of how to use all of them together and fully explore the potential of knowledge graphs in search engines.

This work proposes a new framework for utilizing knowledge bases in information retrieval. Instead of centering around words and using the knowledge graph as an additional resource, this work treats entities equally with words, and represents the query and documents using both word-based and entity-based representations. Thus the interaction of query and document is no longer a `solo' of their words, but a `duet' of their words and entities. 
Working together, the word-based and entity-based representations form a four-way interaction: query words to document words (\texttt{Qw-Dw}), query entities to document words (\texttt{Qe-Dw}), query words to document entities (\texttt{Qw-De}), and query entities to document entities (\texttt{Qe-De}). This leads to a general methodology for incorporating knowledge graphs into text-centric search systems.

The rich and novel ranking evidence from the word-entity duet does come with a cost. Because it is created automatically, the entity-based representation also introduces uncertainties. For example, an entity can be mistakenly annotated to a query, and may mislead the search system.
This paper develops an  attention-based ranking model, \texttt{AttR}-\texttt{Duet}, that employs a simple attention mechanism to handle the noise in the entity representation. The matching component of \texttt{AttR-Duet} focuses on ranking with the word-entity duet, while its attention component focuses on steering the model away from noisy entities.
Trained directly from relevance judgments, \texttt{AttR}-\texttt{Duet} learns how to demote noisy entities and how to rank documents with the word-entity duet simultaneously.

The effectiveness of \texttt{AttR}-\texttt{Duet} is demonstrated on ClueWeb Category B corpora and TREC Web Track queries. On both ClueWeb09-B and ClueWeb12-B13 , \texttt{AttR}-\texttt{Duet} outperforms previous word-based and entity-based ranking systems by at least $14\%$.
We demonstrate that the entities bring additional exact match and soft match ranking signals from the knowledge graph; all entity-based rankings perform similar or better compared to solely word-based rankings. 
We also find that, when the automatically-constructed entity representations are not as clean, the attention mechanism is necessary for the ranking model to utilize the ranking signals from the knowledge graph.
Jointly learned, the attention mechanism is able to demote noisy entities and distill the ranking signals, while without such purification, ranking models become vulnerable to noisy entity representations, and the mixed evidence from the knowledge graph may be more of a distraction than an asset.

In the rest of this paper, Section 2 discusses related work; Section 3 presents the word-entity duet framework for utilizing knowledge graphs in ranking; Section 4 is about the attention-based ranking model;
Experimental settings and evaluations are described in Section 5 and 6; The conclusions and future work are in Section 7.

%% file: Related_work.tex
\section{related work}

There is a long history of research on utilizing semantic resources to improve information retrieval. Controlled vocabulary based information retrieval uses a set of expert-created index terms (mostly organized in ontologies) to represent query and documents. The retrieval is then performed by query and document's overlaps in the controlled vocabulary space, and the human knowledge in the controlled vocabulary is included. Controlled vocabulary is almost a necessity in some special domains. For example, in medical search where queries are often about diseases, treatments, and genes, the search intent may not be covered by the query words, so external information about medical terms is needed. Thesauruses and lexical resources such as WordNet have also been used to address this issue.  Synonyms and related concepts stored in these resources can be added to queries and documents to reduce language varieties and may improve the recall of search results~\cite{li2014semantic}.

Recently, large general domain knowledge bases, such as DBpedia~\cite{dbpedia-swj} and Freebase~\cite{bollacker2008freebase}, have emerged.
Knowledge bases contain human knowledge about real-world entities, such as descriptions, attributes, types, and relationships, usually in the form of knowledge graphs. 
They share the same spirit with controlled vocabularies but are usually created by community efforts or information extraction systems, thus are often at a larger scale.
These knowledge bases provide a new opportunity for search engines to better `understand' queries and documents.  Many new techniques have been developed to explore their potential in various components of ad-hoc retrieval.

An intuitive way is to use the texts associated with related entities to form better query representations. Wikipedia contains well-written entity descriptions and can be used as an external and cleaner pseudo relevance feedback corpus to obtain better expansion terms~\cite{xu2009query}. 
The descriptions of related Freebase entities have been utilized to provide better expansion terms; the related entities are retrieved by entity search~\cite{Jing2016LTR}, or selected from top retrieved documents' annotations~\cite{FACC1}. The text fields of related entities can also be used to provide expanded \textit{learning to rank} features: Entity Query Feature Expansion (EQFE)  expands the query using the texts from related entities' attributes, and these expanded texts generate rich ranking features~\cite{daltonentity}.

Knowledge bases also provide additional connections between query and documents through related entities. 
Latent Entity Space (LES) builds an unsupervised retrieval model that ranks documents based on their textual similarities to latent entities' descriptions~\cite{liu2015latent}. 
EsdRank models the connections between query to entities, and entities to documents using various information retrieval features. These connections are utilized by a latent space learning to rank model, which significantly improved state-of-the-art learning to rank methods~\cite{EsdRank}.

A recent progress is to build entity-based representations for texts.
Bag-of-entities representations built from entity annotations have been used in unsupervised retrieval models, including vector space models~\cite{Xiong2016BOE} and language models~\cite{raviv2016document}.  The entity-based soft match was studied by Semantics-Enabled Language Model (SELM); it connects query and documents in the entity space using their entities' relatedness calculated from an entity linking system~\cite{SELM}.
These unsupervised entity-based retrieval models perform better or can be effectively combined with word-based retrieval models.
Recently, Explicit Semantic Ranking (ESR) performs learning to rank with query and documents' entity representations in scholar search~\cite{ESR}. ESR first trains entity embeddings using a knowledge graph, and then converts the distances in the embedding space to exact match and soft match ranking features, which significantly improved the ranking accuracy of \url{semanticscholar.org}.

%% file: Duet.tex
\section{Word-Entity Duet Framework}
\label{sec:duet}

This section presents our word-entity duet framework for utilizing knowledge bases in search.
Given a query $q$ and a set of candidate documents $D=\{d_1,...,d_{|D|}\}$, our framework aims to provide a systematic approach to better rank $D$ for q, with the help of a knowledge graph (knowledge base) $G$.
In the framework, query and documents are represented by two representations, one word-based and one entity-based (Section~\ref{sec:rep}). The two representations' interactions create the word-entity duet and provide four matching components (Section~\ref{sec:four-way}).

\subsection{Word and Entity Based Representations}
\label{sec:rep}
\textbf{Word-based representations} of query and document are standard bag-of-words:
$\text{Qw}(w) = \text{tf}(w, q)$, and $\text{Dw}(w) = \text{tf}(w, d)$.
Each dimension in the bag-of-words Qw and Dw corresponds to a word $w$. Its weight is the word's frequency (tf) in the query or document. 

A standard approach is to use multiple fields of a document, for example, title and body. Each document field is usually represented by a separate bag-of-words, for example, $\text{Dw}_{\text{title}}$ and $\text{Dw}_{\text{body}}$, and the ranking scores from different fields are combined by ranking models. 
In this work, we assume that a document may have multiple fields. However, to make notation simpler, the field notation is omitted in the rest of this paper unless necessary.

\textbf{Entity-based representations} are bag-of-entities constructed from entity annotations~\cite{Xiong2016BOE}:
$\text{Qe}(e) = \text{tf}(e, q)$ and $\text{De}(e) = \text{tf}(e, d)$, 
where $e$ is an entity linked to the query or the document. We use automatic entity annotations from an entity linking system to construct the bag-of-entities~\cite{Xiong2016BOE}.

An entity linking system finds the entity mentions (surface forms) in a text, and links each surface form to a corresponding entity. For example, the entity `Barack Obama' can be linked to the query `\underline{Obama} Family Tree'. `Obama' is the surface form.

Entity linking systems usually contain two main steps~\cite{TagMe}:
\begin{enumerate}
\item \emph{Spotting}: To find surface forms in the text, for example, to identify the phrase `Obama'.
\item \emph{Disambiguation}: To link the most probable entity from the candidates of each surface form, for example, choosing `Barack Obama' from all possible Obama-related entities.
\end{enumerate}
A commonly used information in \emph{spotting} is the \emph{linked probability} ($lp$), 
the probability of a surface form being annotated in a training corpus, such as Wikipedia. A higher $lp$ means the surface form is more likely to be linked. For example, `Obama' should have a higher $lp$ than `table'. The \emph{disambiguation} usually considers two factors. The first is \emph{commonness} (CMNS), the universal probability of the surface form being linked to the entity. The second is the context in the text, which provides additional evidence for disambiguation. A confidence score is usually assigned to each annotated entity by the entity linking system, based on spotting and disambiguation scores.

The bag-of-entities is not the set of surface forms that appear in the text (otherwise it is not much different from phrase-based representation). Instead, the entities are associated with rich semantics from the knowledge graph. For example, in Freebase, 
the information associated with each entity includes (but is not limited to) its name, alias, type, description, and relations with other entities. The entity-based representation makes these semantics available when matching query and documents.

\begin{table}[t]
\centering
\caption{Ranking features from query words to document words (title and body) ($\Phi_{\texttt{Qw-Dw}}$). \label{tab:qw_dw_feature}}
\begin{tabular}{l|c} 
\hline
Feature Description & Dimension \\ \hline
BM25 & 2 \\
TF-IDF & 2 \\
Boolean OR & 2 \\
Boolean And & 2 \\
Coordinate Match  & 2 \\
Language Model (Lm)  & 2 \\
Lm with JM smoothing & 2 \\
Lm with Dirichlet smoothing & 2 \\
Lm with two-way smoothing & 2 \\ \hline
Total & 18 \\ \hline
\end{tabular}
\end{table}

\begin{table}[t]
\centering
\caption{Ranking features from query entities (name and description) to document words (title and body) ($\Phi_{\texttt{Qe-Dw}}$). \label{tab:qe_dw_feature}}
\begin{tabular}{l|c} 
\hline
Feature Description & Dimension \\ \hline
BM25 & 4 \\
TF-IDF & 4 \\
Boolean Or & 4 \\
Boolean And & 4 \\
Coordinate Match  & 4 \\
Lm with Dirichlet Smoothing & 4 \\ \hline
Total & 24 \\ \hline
\end{tabular}
\end{table}

\subsection{Matching with the Word-Entity Duet}
\label{sec:four-way}
By adding the entity based representation into the search system,
the ranking is no longer a solo match between words, but a word-entity duet that includes four different ways a query can interact with a document: query words to document words (\texttt{Qw-Dw}); query entities to document words (\texttt{Qe-Dw}); query words to document entities (\texttt{Qw-De}); and query entities to document entities (\texttt{Qe-De}). Each of them is a matching component and generates unique ranking features to be used in our ranking model.

\textbf{Query Words to Document Words (\texttt{Qw-Dw}):} This interaction has been widely studied in information retrieval. The matches of Qw and Dw generate term-level statistics such as term frequency and inverse document frequency. These statistics are combined in various ways by standard retrieval models, for example, BM25, language model (Lm), and vector space model. This work applies these standard retrieval models on document title and body fields to extract the ranking features $\Phi_{\texttt{Qw-Dw}}$ in Table~\ref{tab:qw_dw_feature}.

\textbf{Query Entities to Document Words (\texttt{Qe-Dw}):}
Knowledge bases contain textual attributes about entities, such as names and descriptions. These textual attributes make it possible to build cross-space interactions between query entities and document words. Specifically, given a query entity $e$, we use its name and description as pseudo queries, and calculate their retrieval scores on a document's title and body bag-of-words, using standard retrieval models. The retrieval scores from query entities (name or description) to document's words (title or body) are used as ranking features $\Phi_{\texttt{Qe-Dw}}$.
The detailed feature list is in Table~\ref{tab:qe_dw_feature}.

\textbf{Query Words to Document Entities (\texttt{Qw-De}):}
Intuitively, the texts from document entities should help the understanding of the document. For example, when reading a Wikipedia article, the description of a linked entity in the article is helpful for a reader who does not have the background knowledge about the entity. 

The retrieval scores from the query words to document entities' name and descriptions are used to model this interaction.
Different from \texttt{Qe-Dw}, in \texttt{Qw-De}, not all document entities are related to the query.
To exclude unnecessary information, only the highest retrieval scores from each retrieval model are included as features:
\begin{align*}
\Phi_{\texttt{Qw-De}} &\supset \text{max-k}(\{\text{score}(q, e) |\forall e \in \text{De}\}).
\end{align*}
$\text{score}(q, e)$ is the score of q and document entity e from a retrieval model. max-k() takes the k biggest scores from the set. 
Applying retrieval models on title and body entities' names and descriptions, the ranking features $\Phi_{\texttt{Qw-De}}$ in Table~\ref{tab:qw_de_feature} are extracted. We choose a smaller $k$ for title entities as titles are short and rarely have more than three entities.

\begin{table}[t]
\centering
\caption{Ranking features from query words to document entities (name and description) ($\Phi_{\texttt{Qw-De}}$).\label{tab:qw_de_feature}}
\begin{tabular}{l|c} 
\hline
Feature Description & Dimension \\ \hline
Top 3 Coordinate Match on Title Entities & 6 \\
Top 5 Coordinate Match on Body Entities& 10 \\
Top 3 TF-IDF on Title Entities & 6 \\
Top 5 TF-IDF on Body Entities & 10 \\
Top 3 Lm-Dirichlet on Title Entities & 6 \\
Top 5 Lm-Dirichlet on Body Entities & 10 \\ \hline
Total & 48 \\ \hline
\end{tabular}
\end{table}

\begin{table}[t]
\centering
\caption{Ranking features from query entities to document's title and body entities ($\Phi_{\texttt{Qe-De}}$).\label{tab:qe_de_feature}}
\begin{tabular}{l|c} 
\hline
Feature Description & Dimension \\ \hline
Binned translation scores, 1 exact match bin, & \multirow{2}{*}{12} \\
5 soft match Bins in the range $[0, 1)$.  & \\
\hline
\end{tabular}
\end{table}

\textbf{Query Entities to Document Entities (\texttt{Qe-De}):}
There are two ways the interactions in the entity space can be useful. 
The \emph{exact match} signal addresses the vocabulary mismatch of surface forms~\cite{raviv2016document,Xiong2016BOE}. For example, two different surface forms, `Obama' and `US President', are linked to the same entity `Barack Obama' and thus are matched. The \emph{soft match} in the entity space is also useful. For example, a document that frequently mentions `the white house' and `executive order' may be relevant to the query `US President'.

We choose a recent technique, Explicit Semantic Ranking (ESR), to model the exact match and soft match in the entity space~\cite{ESR}.  ESR first calculates an entity translation matrix of the query and document using entity embeddings. Then it gathers ranking features from the matrix by histogram pooling. ESR was originally applied in scholar search and its entity embeddings were trained using domain specific information like author and venue.

In the general domain, there is much research that aims to learn entity embeddings from the knowledge graph~\cite{bordes2013translating, TransR}.
We choose the popular TransE model which is effective and efficient to be applied on large knowledge graphs~\cite{bordes2013translating}.

Given the triples (edges) from the knowledge graph $(e_h, p, e_t)$, including $e_h$ and $e_t$ the head entity and the tail entity, and $p$ the edge type (predicate), TransE learns entity and relationship embeddings ($\vec{e}$ and $\vec{p}$) by optimizing the following pairwise loss:
\begin{align*}
\sum_{(e_h, p, e_t) \in G}\sum_{(e'_h, p, e'_t)  \in G'} [1 + ||\vec{e_h} + \vec{p}- \vec{e_t}||_1 - ||\vec{e'_h} + \vec{p}- \vec{e'_t}||_1]_+,
\end{align*}
where $[\cdot]_+$ is the hinge loss, $G$ is the set of existing edges in the knowledge graph, and $G'$ is the randomly sampled negative instances. 
The loss function ensures that entities in similar graph structures are mapped closely in the embedding space, using the compositional assumption along the edge: $\vec{e_h} + \vec{p} = \vec{e_t}$. 

The distance between two entity embeddings describes their similarity in the knowledge graph~\cite{bordes2013translating}. Using L1 similarity, 
a translation matrix can be calculated:
\begin{align}
T(e_i, e_j)  &= 1 - ||\vec{e}_i - \vec{e}_j||_1.
\end{align}
$T$ is the $|\text{Qe}| \times |\text{De}|$ translation matrix. $e_i$ and $e_j$ are the entities in the query and the document respectively. 

Then the histogram pooling technique is used to gather query-document matching signals from $T$~\cite{jiafeng2016deep, ESR}:
\begin{align*}
& \vec{S}(\text{De}) = \max_{e \in \text{Qe}} T(e, \text{De}) \\
& B_k(\vec{S}(\text{De})) = \log \sum_j I(st_k \leq \vec{S}_j(\text{De}) < ed_k).
\end{align*}
$\vec{S}(d)$ is the max-pooled $|\text{De}|$ dimensional vector, whose $j^{th}$ dimension is the maximum similarity of the $j^{th}$ document entity to any query entities. $B_k()$ is the $k^{th}$ bin that counts the number of translation scores in its range $[st_k, ed_k)$. 

We use the same six bins as in the ESR paper: $[1, 1],$ $[0.8, 1),$ $[0.6, 0.8),$ $[0.4, 0.6),$ $[0,2, 0.4),$ $[0, 0,2)$.
The first bin is the exact match bin and is equivalent to the entity frequency model~\cite{Xiong2016BOE}. The other bin scores capture the soft match signal between query and documents at different levels. These bin scores generated the ranking features $\Phi_{\texttt{Qe-De}}$ in Table~\ref{tab:qe_de_feature}.

\subsection{Summary\label{sec:special_case}}
The word-entity duet incorporates various semantics from the knowledge graph: The textual attributes of entities are used to model the cross-space interactions (\texttt{Qe-Dw} and \texttt{Qw-De}); the relations in the knowledge graphs are used to model the interactions in the entity space (\texttt{Qe-De}), through the knowledge graph embedding. The word-based retrieval models are also included (\texttt{Qw-Dw}).

Many prior methods are included in the duet framework. 
For example, the query expansion methods using Wikipedia or Freebase represent the query using related entities, and then use these entities' texts to build additional connections with the document's text~\cite{xu2009query,xiong2015fbexpansion,daltonentity}; the latent entity space  techniques first find a set of highly related query entities, and then rank documents using their connections with these entities~\cite{liu2015latent,EsdRank}; the entity based ranking methods model the interactions between query and documents in the entity space using exact match~\cite{raviv2016document,Xiong2016BOE} and soft match~\cite{ESR}. 

Each of the four interactions generates a set of ranking signals. A straightforward way is to use them as features in learning to rank models. However, the entity representations may include noise and generate misleading ranking signals, which motivates our \texttt{AttR-Duet} ranking model in the next section.

%% file: Att_ltr.tex
\section{Attention Based Ranking Model}

\begin{figure*}
\centering
\includegraphics[width=1\textwidth]{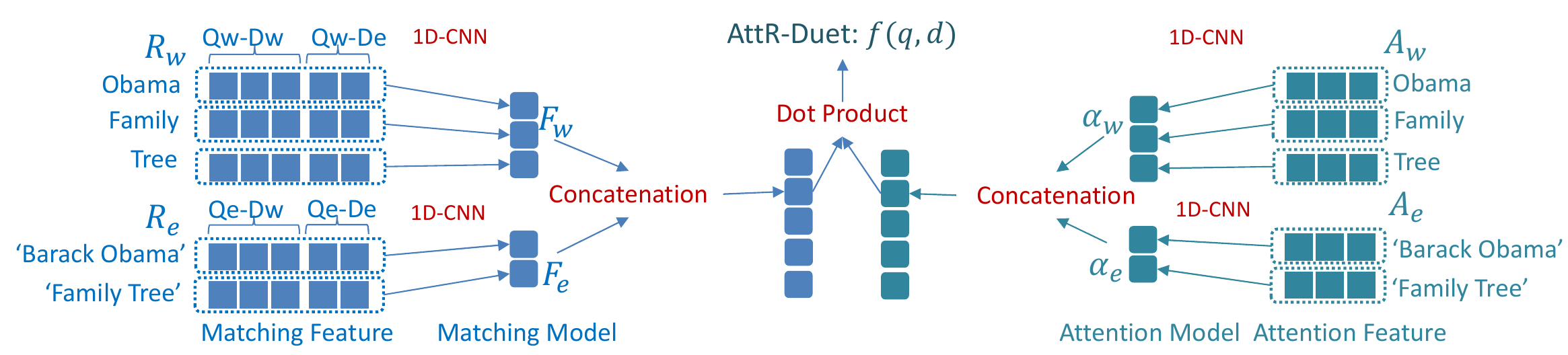}
\caption{The Architecture of the Attention based Ranking Model for Word-Entity Duet (\texttt{AttR-Duet)}. 
The left side models the query-document matching in the word-entity duet. The right side models the importances of query entities using attention features. They together produce the final ranking score.
\label{fig:att}}
\end{figure*}

\begin{table}[t]
\centering
\caption{Attention features for query entities.\label{tab:att_feature}}
\begin{tabular}{l|c} \hline
Feature Description & Dimension \\ \hline
Entropy of the Surface Form & 1 \\
Is the Most Popular Candidate Entity  & 1 \\
Margin to the Next Candidate Entity & 1 \\
Embedding Similarity with Query& 1\\
\hline 
Total & 4 \\ \hline
\end{tabular}
\end{table}

Unlike bag-of-words, entity-based representations are constructed using automatic entity linking systems. 
It is inevitable that some entities are mistakenly annotated, especially in short queries where there is less context for disambiguation. If an unrelated entity is annotated to the query, it will introduce misleading ranking features; documents that match the unrelated entity might be promoted. Without additional information, ranking models have little leverage to distinguish the useful signals brought in by correct entities from those by the noisy ones, and their accuracies might be limited.

We address this problem with an attention based ranking model \texttt{AttR-Duet}.
It first extracts attention features to describe the quality of query entities.
Then \texttt{AttR-Duet} builds a simple attention mechanism using these features to demote noisy entities. The attention and the matching of query-documents are trained
together using back-propagation, enabling the model to learn simultaneously how to weight entities of varying quality and how to rank with the word-entity duet. 
The attention features are described in Section~\ref{sec:att_feature}. The details of the ranking model are discussed in Section~\ref{sec:model}.

\subsection{Attention Features}
\label{sec:att_feature}
Two groups of attention features are extracted for each query entity
to model its annotation ambiguity and its closeness to the query.

\textbf{Annotation Ambiguity} features describe the \textit{risk} of an entity annotation.
There is a risk that the linker may fail to disambiguate the surface form to the correct entity, especially when the surface form is too ambiguous. For example, `Apple' in a short query can be the fruit or the brand. It is risky to put high attention on it.
There are three ambiguity features used in \texttt{AttR-Duet}.

The first feature is the entropy of the surface form. Given a training corpus, for example, Wikipedia, we gather the probability of a surface form being linked to different candidate entities, and calculate the entropy of this probability. 
The higher the entropy is, the more ambiguous the surface form is, and the less attention the model should put on the corresponding query entity.
The second feature is whether the annotated entity is the most popular candidate of the surface form, i.e.~has the highest commonness score (CMNS). 
The third feature is the difference between the linked entity's CMNS to the next candidate entity's. 

A \textbf{closeness} attention feature is extracted using the distance between the query entity and the query words in an embedding space.
An entity and word joint embedding model are trained on a corpus including the original documents and the documents with surface forms replaced by linked entities. The cosine similarity between the entity embedding to the query embedding (the average of its words' embeddings) is used as the feature.
Intuitively, a higher similarity score should lead to more attention.

The full list of entity attention features, $\text{Att}(e)$, is listed in Table~\ref{tab:att_feature}.

\subsection{Model}
\label{sec:model}

The architecture of \texttt{AttR-Duet} is illustrated in Figure~\ref{fig:att}. 
It produces a ranking function $f(q, d)$ that re-ranks candidate documents $D$ for the query $q$, with the ranking features in Table~\ref{tab:qw_dw_feature}-\ref{tab:qe_de_feature} and attention features in Table~\ref{tab:att_feature}. $f(q, d)$ is expected to weight query elements more properly and rank document more accurately.

\textbf{Model inputs:}
Suppose the query contains words $\{w_1,...,w_n\}$ and entities $\{e_1,$ $...$ $,e_m\}$, there are four input feature matrices: $R_w$, $R_e$, $A_w$, and $A_e$. $R_w$ and $R_e$ are the ranking feature matrices for query words and entities in the document. $A_w$ and $A_e$ are the attention feature matrices for words and entities.
These matrices' rows are feature vectors previously described:
\begin{align}
R_w(w_i, \cdot) &= \Phi_{\texttt{Qw-Dw}}(w_i) \sqcup \Phi_{\texttt{Qw-De}}(w_i) \\
R_e(e_j, \cdot) &= \Phi_{\texttt{Qe-Dw}}(e_j) \sqcup \Phi_{\texttt{Qe-De}}(e_j) \\
A_w(w_i, \cdot) &= 1 \\
A_e(e_j, \cdot) &= \text{Att}(e_j).
\end{align}
$\Phi_{\texttt{Qw-Dw}}$, $\Phi_{\texttt{Qw-De}}$, $\Phi_{\texttt{Qe-Dw}}$, and $\Phi_{\texttt{Qe-De}}$ are the ranking features from the word-entity duet, as described in Section~\ref{sec:duet}. $\sqcup$ concatenates the two feature vectors of a query element.
$\text{Att}(e_j)$ is the attention features for entity $e_j$ (Table~\ref{tab:att_feature}). 
In this work, we use uniform word attention ($A_w=1$), because the main goal of the attention mechanism is to handle the uncertainty in the entity representations.

The \textbf{matching part} contains two Convolutional Neural Networks (CNN's). One matches query words to $d$ ($R_w$); the other one matches query entities to $d$ ($R_e$). The convolution is applied on the query element (word/entity) dimension, assuming that the ranking evidence from different query words or entities should be treated the same.
The simplest setup with one 1d CNN layer, 1 filter, and linear activation function can be considered as a linear model applied `convolutionally' on each word or entity:
\begin{align}
F_w(w_i) &= W^m_w \boldsymbol{\cdot} R_w(w_i, \cdot) + b^m_w \\
F_e(e_j) &= W^m_e \boldsymbol{\cdot} R_e(e_j, \cdot) + b^m_e.
\end{align}
$F_w(w_i)$ and $F_e(e_j)$ are the matching scores from query word $w_i$ and query entity $e_j$, respectively. The matching scores from all query words form an $n$ dimensional vector $F_w$, and those from entities form an $m$ dimensional vector $F_e$. $W^m_w, W^m_e, b^m_w$, and $b^m_e$ are the matching parameters to learn.

The \textbf{attention part} also contains two CNN's. One weights query words with $A_w$ and the other one weights query entities with $A_e$. The same convolution idea is applied as the attention features on each query word/entity should be treated the same.

The simplest set-ups with one CNN layer are:
\begin{align}
\alpha_w(w_i) &= \text{ReLU}(W^a_w \boldsymbol{\cdot} A_w(w_i, :) + b^a_w) \\
\alpha_e(e_j) &= \text{ReLU}(W^a_e \boldsymbol{\cdot} A_e(e_j, :) + b^a_e).
\end{align}
$\alpha_w(w_i)$ and $\alpha_e(e_j)$ are the attention weights on word $w_i$ and entity $e_j$.  $\{W^a_w, W^a_e, b^a_w, b^a_e\}$ are the attention parameters to learn. 
ReLU activation is used to ensure non-negative attention weights. The matching scores can be negative because only the differences between documents' matching scores matter.

The final ranking score combines the matching scores using the attention scores:
\begin{align}
f(q, d) &= F_w \boldsymbol{\cdot} \alpha_w + F_e \boldsymbol{\cdot} \alpha_e.
\end{align}

The \textbf{training} is done by optimizing the pairwise hinge loss:
\begin{align}
l(q, D) &= \sum_{d \in D^+} \sum_{d' \in D^-} [1 - f(q, d) + f(q, d')]_+.
\end{align}
$D^+$ and $D^-$ are the set of relevant documents and the set of irrelevant documents. $[\cdot]_+$ is the hinge loss. The loss function can be optimized using back-propagation in the neural network, and the matching part and the attention part are learned simultaneously.

%% file: Experiment.tex
\section{Experimental Methodology}

This section describes the experiment methodology, including dataset, baselines,  and the implementation details of our methods.

\textbf{Dataset:} Ranking performances were evaluated on the TREC Web Track ad-hoc task, the standard benchmark for web search. TREC 2009-2012 provided 200 queries for ClueWeb09, and TREC 2013-2014 provided 100 queries for ClueWeb12. The Category B of both corpora (ClueWeb09-B and ClueWeb12-B13) and corresponding TREC relevance judgments were used. 

On ClueWeb09-B, the SDM runs provided by EQFE~\cite{daltonentity} are used as the base retrieval.
It is a well-tuned Galago-based implementation and performs better than Indri's SDM. All their settings are inherited, including spam filtering using waterloo spam score (with a threshold of 60), INQUERY plus web-specific stopwords removal, and KStemming.
On ClueWeb12-B13, not all queries' rankings are available from prior work, and Indri's SDM performs similarly to language model. For simplicity, the base retrieval on ClueWeb12-B13 used is Indri's default language model with KStemming, INQUERY stopword removal, and no spam filtering. All our methods and learning to rank baselines re-ranked the first 100 documents from the base retrieval. 

The ClueWeb web pages were parsed using Boilerpipe\footnote{https://github.com/kohlschutter/boilerpipe}. The `KeepEverythingExtractor' was used to keep as much text from the web page as possible, to minimize the parser's influence. The documents were parsed to two fields: title and body.
All the baselines and methods implemented by ourselves were built upon the same parsed results for fair comparisons.

The \textbf{Knowledge Graph} used in this work is Freebase~\cite{bollacker2008freebase}. 
The query and document entities were both annotated by TagMe~\cite{TagMe}. 
No filter was applied on TagMe's results; all annotation were kept.
This is the most widely used setting of entity-based ranking methods on ClueWeb~\cite{raviv2016document, EsdRank, Xiong2016BOE}.

\textbf{Baselines} included standard word-based baselines: Indri's language model (\texttt{Lm}), sequential dependency model (\texttt{SDM}), and two state-of-the-art learning to rank methods:
\texttt{RankSVM}\footnote{https://www.cs.cornell.edu/people/tj/svm\_light/svm\_rank.html}~\cite{ranksvm} and coordinate ascent (\texttt{Coor-Ascent}\footnote{https://sourceforge.net/p/lemur/wiki/RankLib/})~\cite{metzler2007linear}.
\texttt{RankSVM} was trained and evaluated using a 10-fold cross-validation on each corpus. Each fold was split to train (80\%), develop (10\%), and test (10\%). The develop part was used to tune the hyper-parameter c of the linear SVM from the set $\{1e-05, 0.0001,$ $0.001, 0.01,$ $0.03, 0.05,$ $0.07, 0.1,$ $0.5, 1\}$. \texttt{Coor-Ascent} was trained using RankLib's recommended settings, which worked  well in our experiments. They used the same word based ranking features as in Table~\ref{tab:qw_dw_feature}. 

Entity-based ranking baselines were also compared.
\texttt{EQFE}~\cite{daltonentity}, \texttt{EsdRank}~\cite{EsdRank}, and \texttt{BOE-TagMe}~\cite{Xiong2016BOE} runs are provided on their authors' websites.
The comparisons with \texttt{EQFE} and \texttt{EsdRank} were mainly done on ClueWeb09 as the full ranking results on ClueWeb12 are not publicly available. \texttt{BOE-TagMe} is the best TagMe based runs, which is TagMe-EF on ClueWeb09-B and TagMe-COOR on ClueWeb12-B13~\cite{Xiong2016BOE}.
Explicit Semantic Ranking (ESR) was implemented by ourselves as originally it was only applied on scholar search~\cite{ESR}.

There are also other \textit{unsupervised} entity-based systems~\cite{liu2015latent, raviv2016document, xiong2015fbexpansion}; it is unfair to compare them with supervised methods.

\textbf{Evaluation} was done by NDCG@20 and ERR@20, the official TREC Web Track ad-hoc task evaluation metrics. Statistical significances were tested by
permutation test with p$ <0.05$.

\textbf{Feature Details:} 
All parameters in the unsupervised retrieval model features were kept default. All texts were reduced to lower case, punctuation was discarded, and standard INQUERY stopwords were removed.  Document fields included title and body, both parsed by Boilerpipe. Entity textual fields included name and description. 
When extracting \texttt{Qw-De} features, if a document did not have enough entities (3 in title or 5 in body), the feature values were set to $-20$. The 
TransE embeddings were trained using Fast-TransX library\footnote{https://github.com/thunlp/Fast-TransX}. The embedding dimension used is 50.  

When extracting the attention features in Table~\ref{tab:att_feature},
the word and entity joint embeddings were obtained by training a skip-gram model on the candidate documents using Google's word2vec~\cite{word2vec} with 300 dimensions;
the surface form's statistics were calculated from Google's FACC1 annotation~\cite{FACC1}.

\textbf{Model Details:} \texttt{AttR-Duet} was evaluated using 10-fold cross validation with the same partitions as \texttt{RankSVM}. Deeper neural networks were explored but did not provide much improvement so the simplest CNN setting was used: 1 layer, 1 filter, linear activation for the ranking part, and ReLU activation for the attention part.
All CNN's weights were L2-regularized. Regularization weights were selected from the set $\{0, 0.001, 0.01, 0.1\}$ using the develop fold in the cross validation. 
Training loss was optimized using the Nadam algorithm~\cite{sutskever2013importance}. 
Our implementation was based on Keras. To facilitate implementation, input feature matrices of query elements were padded to the maximum length with zeros. 
Batch training was used, given the small size of training data.

Using a common CPU, the training took 4-8 hours to converge on ClueWeb09-B and 2-4 hours on ClueWeb12-B13. 
The testing is efficient as the neural network is shallow. 
The document annotations, TransE embeddings, and surface form information can be obtained off line. Query entity linking is efficient given the short query length. If the embedding results and entities' texts are maintained in memory, the feature extraction is of the same complexity as typical learning to rank features.

The rankings, evaluation results, and the data used in our experiments are available online at \url{http://boston.lti.cs.cmu.edu/appendices/SIGIR2017_word_entity_duet/}.

%% file: Evaluation.tex
\begin{table*}[t]
\centering
\caption{Overall accuracies of \texttt{AttR-Duet} and baselines. 
(U) and (S) indicate unsupervised or supervised method. (E) indicates that information from entities is used.
Relative performances compared with \texttt{RankSVM} are shown in percentages. \textbf{W}in/\textbf{T}ie/\textbf{L}oss are the number of queries a method improves, does not change, or hurts, compared with \texttt{RankSVM} on NDCG@20. Best results in each metric are marked \textbf{bold}. 
$\mathsection$ marks
statistically significant improvements (p$<0.05$) over \emph{all} baselines.
\label{tab:overall}
}

\begin{tabular}{lc|lr|lr|c||lr|lr|c}
\hline
&
 & \multicolumn{5}{c||}{\bf{ClueWeb09-B}}
 & \multicolumn{5}{c}{\bf{ClueWeb12-B13}}
 \\ \hline 
\multicolumn{2}{c|}{\bf{Method}} &
\multicolumn{2}{c|}{\bf{NDCG@20}} &
\multicolumn{2}{c|}{\bf{ERR@20}} &
\bf{W/T/L} &
\multicolumn{2}{c|}{\bf{NDCG@20}} &
\multicolumn{2}{c|}{\bf{ERR@20}} &
\bf{W/T/L}
\\ \hline

\texttt{Lm} & (U)
 & ${0.1757}$ & $ -33.33\%  $
 & ${0.1195}$ & $ -22.63\%  $
& 47/28/125

 & ${0.1060}$ & $ -12.02\%  $
& ${0.0863}$ & $ -6.67\%  $
& 35/22/43
\\
\texttt{SDM} & (U)
 & ${0.2496}$ & $ -5.26\%  $
 & ${0.1387}$ & $ -10.20\%  $
& 62/38/100
 & ${0.1083}$ & $ -10.14\%  $
 & ${0.0905}$ & $ -2.08\%  $
& 27/25/48

\\ \hline

\texttt{RankSVM} & (S)
 & $0.2635$ & --  &  $0.1544$ & --  & --/--/--
 
  & $0.1205$ & --  &  $0.0924$ & --  & --/--/--
 \\
\texttt{Coor-Ascent} & (S)
 & ${0.2681}$ & $ +1.75\%  $

 & ${0.1617}$ & $ +4.72\%  $

& 71/47/82
 & ${0.1206}$ & $ +0.08\%  $
 & ${0.0947}$ & $ +2.42\%  $
& 36/32/32

\\ \hline
\texttt{BOE-TagMe} & (UE)
 & ${0.2294}$ & $ -12.94\%  $
 & ${0.1488}$ & $ -3.63\%  $
& 74/25/101
& ${0.1173}$ & $ -2.64\%  $
& ${0.0950}$ & $ +2.83\%  $
& 44/19/37
\\ \hline

\texttt{ESR} & (SE)
 & ${0.2695}$ & $ +2.30\%  $
 & ${0.1607}$ & $ +4.06\%  $
& 80/39/81 
& ${0.1166}$ & $ -3.22\%  $
 & ${0.0898}$ & $ -2.81\%  $
& 30/23/47 \\

\texttt{EQFE}\footnotemark[5] & (SE)
 & ${0.2448}$ & $ -7.10\%  $
 & ${0.1419}$ & $ -8.10\%  $
& 77/33/90
& n/a & -- & n/a & -- & --/--/--
\\

\texttt{EsdRank}\footnotemark[5] & (SE)
 & ${0.2644}$ & $ +0.33\%  $
 & ${0.1756}$ & $ +13.69\%  $
& 88/28/84
& n/a & -- & n/a & -- & --/--/--

\\ \hline
\texttt{AttR-Duet} & (SE)
 & $\bf{0.3197}^{\mathsection}$ & $ +21.32\%  $

 & $\bf{0.2026}^{\mathsection}$ & $ +31.21\%  $

& 101/37/62

& $\bf{0.1376}^{\mathsection}$ & $ +14.22\%  $

 & $\bf{0.1154}^{\mathsection}$ & $ +24.92\%  $

& 45/24/31

\\
 \hline

\end{tabular}
\end{table*}

\begin{table*}[th]
\centering
\caption{Ranking accuracy with each group of matching feature from the word-entity duet. 
\texttt{Base Retrieval} is \texttt{SDM} on ClueWeb09 and \texttt{Lm} on ClueWeb12.
\texttt{LeToR-Qw-Dw} uses the query and document's BOW (Table~\ref{tab:qw_dw_feature}). \texttt{LeToR-Qe-Dw} uses the query's BOE and document's BOW (Table~\ref{tab:qe_dw_feature}), \texttt{LeToR-Qw-De} is the query BOW + document BOE (Table~\ref{tab:qw_de_feature}), and \texttt{LeToR-Qe-De} uses the query and document's BOE (Table~\ref{tab:qe_de_feature}). \texttt{LeToR-All} uses all groups.
Relative performances in percentages, \textbf{W}in/\textbf{T}ie/\textbf{L}oss on NDCG@20, and statistically significant improvements ($\dagger$) are all compared with \texttt{Base Retrieval}.
\label{tab.ranking_res}
}

\begin{tabular}{l|lr|lr|c||lr|lr|c}
\hline
 & \multicolumn{5}{c||}{\bf{ClueWeb09-B}}
 & \multicolumn{5}{c}{\bf{ClueWeb12-B13}}
 \\ \hline 
\multicolumn{1}{c|}{\bf{Method}} &
\multicolumn{2}{c|}{\bf{NDCG@20}} &
\multicolumn{2}{c|}{\bf{ERR@20}} &
\bf{W/T/L} &
\multicolumn{2}{c|}{\bf{NDCG@20}} &
\multicolumn{2}{c|}{\bf{ERR@20}} &
\bf{W/T/L}
\\ \hline

\texttt{Base Retrieval}
 & ${0.2496}$ & $ --  $
 & ${0.1387}$ & $ -- $
& --/--/--
 & ${0.1060}$ & $ --  $
& ${0.0863}$ & $ --  $
& --/--/--
\\ \hline

\texttt{LeToR-Qw-Dw} 
 & ${0.2635}^{\dagger }$ & $ +5.55\%  $

 & ${0.1544}^{\dagger }$ & $ +11.36\%  $

& 100/38/62

 & $\bf{0.1205}$ & $ +13.67\%  $

 & ${0.0924}$ & $ +7.14\%  $

& 43/22/35
\\ 

\texttt{LeToR-Qe-Dw}
 & ${0.2729}$ & $ +9.33\%  $

 & $\bf{0.1824}^{\dagger  }$ & $ +31.51\%  $

& 82/34/84

 & ${0.1110}$ & $ +4.66\%  $

 & $\bf{0.0928}$ & $ +7.63\%  $

& 40/20/40

\\
\texttt{LeToR-Qw-De}
 & $\bf{0.2867}^{\dagger }$ & $ +14.83\%  $

 & ${0.1651}^{\dagger }$ & $ +19.07\%  $

& 91/39/70

 & ${0.1146}$ & $ +8.09\%  $

 & ${0.0880}$ & $ +1.96\%  $

& 42/20/38

\\
\texttt{LeToR-Qe-De} 
 & ${0.2695}^{\dagger }$ & $ +7.97\%  $

 & ${0.1607}^{\dagger }$ & $ +15.88\%  $

& 99/40/61

 & ${0.1166}$ & $ +10.01\%  $

 & ${0.0898}$ & $ +4.13\%  $

& 38/20/42

\\ \hline
\texttt{LeToR-All}
 & $\bf{0.3099}^{\dagger  }$ & $ +24.13\%  $

 & $\bf{0.1955}^{\dagger  }$ & $ +40.97\%  $

& 103/38/59

 & $\bf{0.1205}$ & $ +13.69\%  $

 & $\bf{0.1000}$ & $ +15.93\%  $

& 47/19/34
\\ \hline
\end{tabular}
\end{table*}

\section{Evaluation Results}
This section first evaluates the overall ranking performances of the word-entity duet with attention based learning to rank. 
Then it analyzes the two parts of \texttt{AttR-Duet}: Matching with the word-entity duet and the attention mechanism.

\subsection{Overall Performance}

The overall accuracies of  \texttt{AttR-Duet} and baselines are shown in Table~\ref{tab:overall}. 
Relative performances over \texttt{RankSVM} are shown in percentages.
\textbf{W}in/\textbf{T}ie/\textbf{L}oss are the number of queries a method improves, does not change, and hurts, compared with \texttt{RankSVM} on NDCG@20. Best results in each metric are marked \textbf{Bold}. $\mathsection$ indicates statistically significant improvements over \emph{all} available baselines.

\texttt{AttR-Duet} outperformed all baselines significantly by large margins. On ClueWeb09-B, a widely studied benchmark for web search, \texttt{AttR-Duet} improved \texttt{RankSVM}, a strong learning to rank baseline, by more than $20\%$ at NDCG@20, and more than $30\%$ at ERR@20, showing the advantage of the word-entity duet over bag-of-words. 
\texttt{ESR}, \texttt{EQFE} and \texttt{EsdRank}, previous state-of-the-art entity-based ranking methods, were outperformed by at least $15\%$. 
It is not surprising because the word-entity duet framework was designed to include all of their effects, as discussed at Section~\ref{sec:special_case}.
ClueWeb12-B13 has been considered a hard dataset due to its noisy corpus and harder queries. The size of its training data is also smaller, which limits the strength of our neural network. However, \texttt{AttR-Duet} still significantly outperformed all available baselines by at least $14\%$. The information from entities is effective and also different with those from words: \texttt{AttR-Duet} influences more than three-quarters of the queries, and improves the majority of them.

\addtocounter{footnote}{1}
\footnotetext{Ranking results are obtained from the authors' websites. ClueWeb09-B scores are higher than in original papers~\cite{daltonentity,EsdRank} as we evaluate them using TREC's Category B qrels. 
The original papers used Category A's qrels although they ranked Category B documents.
EQFE and EsdRank's ClueWeb12 results are not available as they were only evaluated on the first 50 queries of the 100.}

\subsection{Matching with Word-Entity Duet~\label{sec.rank_eva}}

\begin{table*}
\centering
\caption{Examples of entities used in \texttt{Qw-De} and \texttt{Qe-De}.
The first half are examples of matched entities in relevant and irrelevant documents, which are used to extract \texttt{Qw-De} features. The second half are examples of entities falls into the exact match bin and the closest soft match bins, used to extract \texttt{Qe-De} features.
\label{tab:eg_feature}
}
\begin{tabular}{l|c|c} 
\hline
\multicolumn{3}{c}{\textbf{Examples of Most Similar Entities to the Query} } \\ \hline
\bf{Query} & \bf{Top Entities in Relevant Documents} & \bf{Top Entities in Irrelevant Documents} 
% & \bf{NDCG Gain}
\\ \hline
Uss Yorktown Charleston SC & `USS Yorktown (CV-10)' & `Charles Cornwallis', `USS Yorktown (CV-5)' 
% & $0.3040$
\\ \hline
Flushing & `Roosevelt Avenue', `Flushing, Queens' &  `Flushing (physiology)',  `Flush (cards)'
%  & $0.6923$ 
 \\
\hline 
\hline
\multicolumn{3}{c}{\textbf{Examples of Neighbors in Knowledge Graph Embedding} } \\  \hline
\bf{Query} & \bf{Entities in Exact Match Bin} & \bf{Entities in Soft Match Bins}
% & NDCG Gain
\\ \hline
Uss Yorktown Charleston SC & `USS Yorktown (CV-10)', `Charleston, SC'
& `Empire of Japan',  `World War II'   
% & $0.1432$ 
\\ \hline
Flushing &  `Flushing, Queens' &  `Brooklyn', `Manhattan', `New York'
%  & $0.2085$ 
 \\
\hline
\end{tabular}
\end{table*}

\begin{figure*}[t]
\centering
\begin{subfigure}{0.47\linewidth}
\centering
\includegraphics[width=1\linewidth]{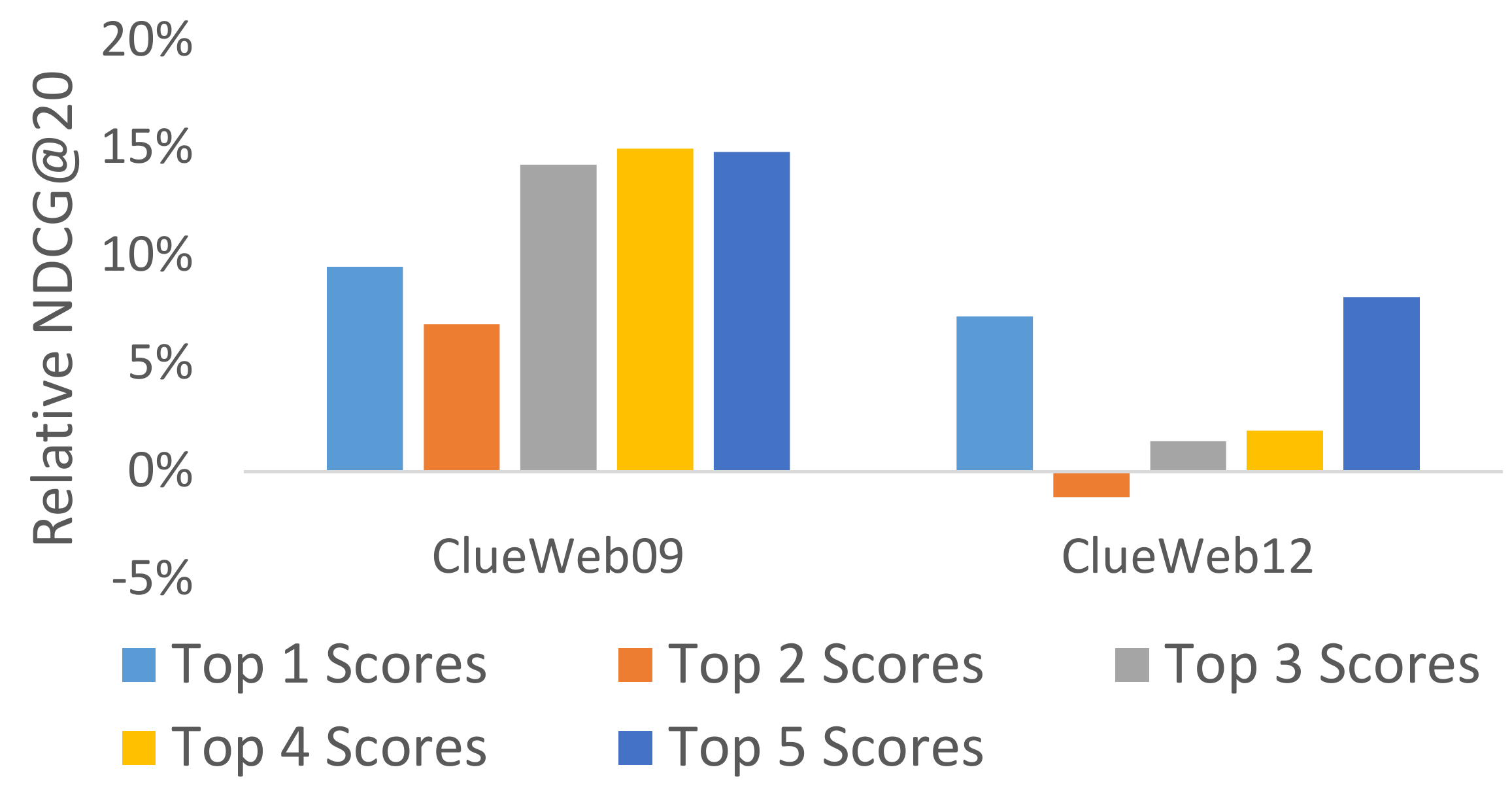}
\caption{Features from Query Words to Document Entities (\texttt{Qw-De})
\label{fig:dexp_eva}}
\end{subfigure}
% \vspace{0.1cm}
\begin{subfigure}{0.45\linewidth}
\centering
\includegraphics[width=1\linewidth]{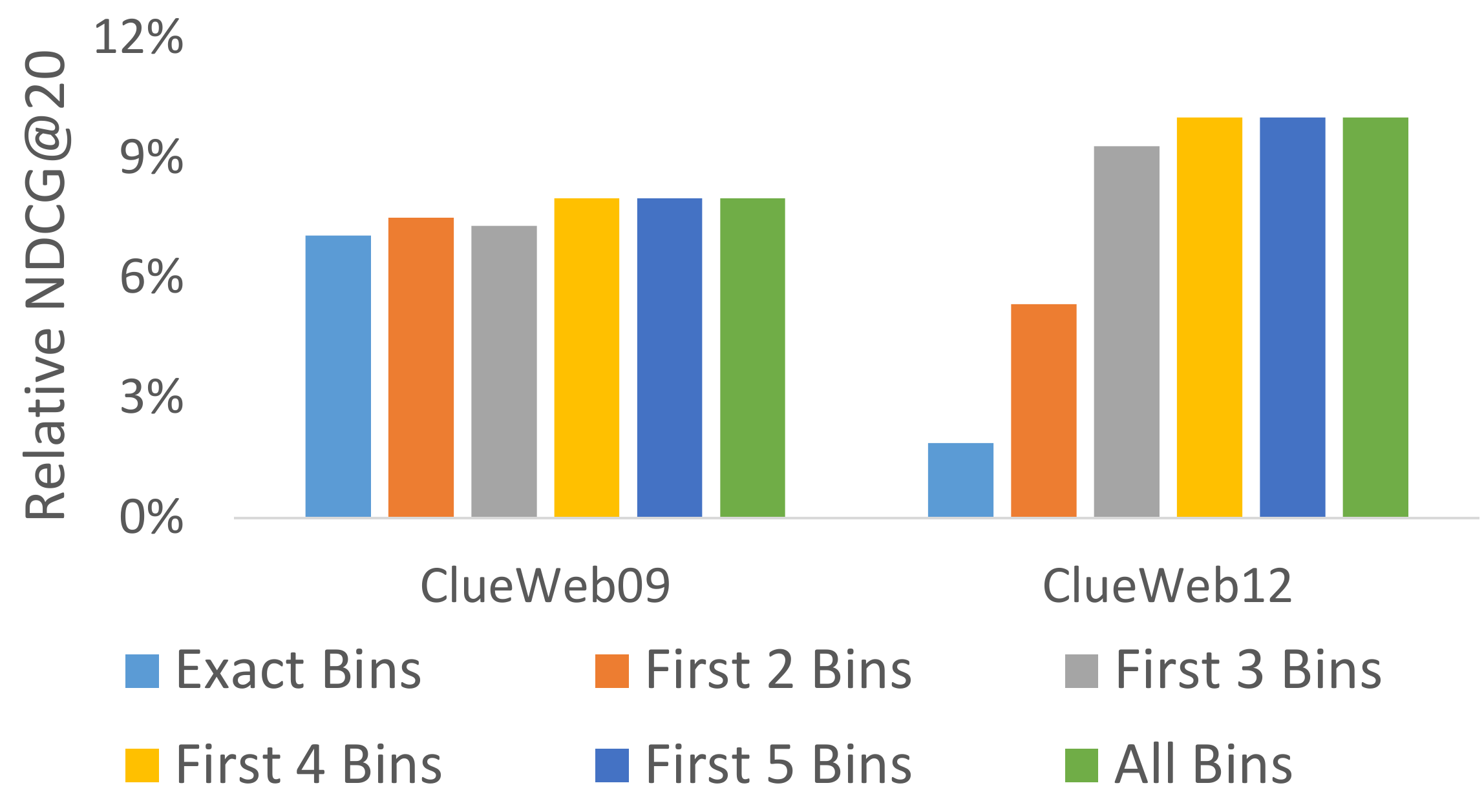}
\caption{Features from Query Entities to Document Entities (\texttt{Qe-De})
\label{fig:esr_eva}}
\end{subfigure}
\caption{Incremental ranking feature analysis.
The y-axis is the relative NDCG@20 improvement over the base retrieval. The x-axis refers to the features from only top k (1-5) entity match scores (\ref{fig:dexp_eva}), and the features from only first k (1-6) bins in the \texttt{ESR} model (\ref{fig:esr_eva}), both ordered incrementally from left to right.}
\end{figure*}

In a sophisticated system like \texttt{AttR-Duet}, it is hard to tell the contributions of different components. 
This experiment studies how each of the four-way interactions in the word-entity duet contributes to the ranking performance individually.
For each group of the matching features in Table~\ref{tab:qw_dw_feature}-~\ref{tab:qe_de_feature}, we train a RankSVM individually, which resulted in four ranking models: \texttt{LeToR-Qw-Dw}, \texttt{LeToR-Qe-Dw}, \texttt{LeToR-Qw-De}, and \texttt{LeToR-Qe-De}. \texttt{LeToR-All} which uses all ranking features is also evaluated.
% All of these methods use the same setting with \texttt{RankSVM}. 
In \texttt{LeToR-Qw-De} and \texttt{LeToR-Qe-De}, the score of the base retrieval model is included as a feature, so that there is a feature to indicate the strength of the word-based match for the whole document. All these methods were trained and tested in the same setting as \texttt{RankSVM}. As a result, \texttt{LeToR-Qw-Dw} is equivalent to the \texttt{RankSVM} baseline, and \texttt{LeToR-Qe-De} is equivalent to the \texttt{ESR} baseline.

Their evaluation results are listed in Table~\ref{tab.ranking_res}. 
Relative performances (percentages), \textbf{W}in/\textbf{T}ie/\textbf{L}oss, and statistically significant improvements ($\dagger$) are all compared with \texttt{Base Retrieval} (\texttt{SDM} on ClueWeb09 and \texttt{Lm} on ClueWeb12). 
All four groups of matching features were able to improve the ranking accuracy of \texttt{Base Retrieval} when used individually as ranking features, demonstrating the usefulness of all matching components in the duet.
On ClueWeb09-B, all three entity related components, \texttt{LeToR-Qe-Dw}, \texttt{LeToR-Qw-De}, and \texttt{LeToR-Qe-De}, provided similar or better performances than the word-based \texttt{RankSVM}. When all features were used together, \texttt{LeToR-All} significantly improved \texttt{RankSVM} by $17\%$ and $26\%$ on NDCG@20 and ERR@20, showing that the ranking evidence from different parts of the duet can reinforce each other.  

On ClueWeb12-B13, entity-based matching was less effective. \texttt{LeToR-All}'s NDCG@20 was the same as \texttt{RankSVM}'s, despite additional matching features. The difference is that the annotation quality of TagMe on ClueWeb12 queries is lower (Table~\ref{tab:att_gain})~\cite{Xiong2016BOE}. The noisy entity representation may mislead the ranking model, and prevent the effective usage of entities. To deal with this uncertainty is the motivation of the attention based ranking model, which is studied in Section~\ref{sec:att_eva}.

\subsection{Matching Feature Analysis}
The features from the word space (\texttt{Qw-Dw}) are well understood, and the feature from the query entities to document words (\texttt{Qe-Dw}) have been studied in prior research~\cite{daltonentity,liu2015latent,EsdRank}. 
This experiment analyzes the features from the two new components (\texttt{Qw-De} and \texttt{Qe-De}).

\textbf{\texttt{Qw-De} features} match the query words with the document entities.  For each document, the query words are matched with the textual fields of document entities using retrieval models, and the highest scores are \texttt{Qw-De} features. 

We performed an incremental feature analysis of \texttt{LeToR-Qw-De}. Starting with the highest scored entities from each group in Table~\ref{tab:qw_de_feature}, we incrementally added the next highest ones to the model and evaluated the ranking performance.
The results are shown in Figure~\ref{fig:dexp_eva}. The y-axis is the relative NDCG@20 improvements over the base retrieval model. The x-axis is the used features. For example, `Top 3 Scores' uses the top 3 entities' retrieval scores in each row of Table~\ref{tab:qw_de_feature}.

The highest scores were very useful. Simply combining them with the base retrieval provided nearly 10\% gain on ClueWeb09-B and about 7\% on ClueWeb12-B13. 
Adding the following scores was not that stable, perhaps because the corresponding entities were rather noisy, given the simple retrieval models used to match query words with them.
Nevertheless, the top 5 scores together further improve the ranking accuracy.

The first half of Table~\ref{tab:eg_feature} shows examples of entities with highest matching scores. We found that such `top' entities from relevant documents are frequently related to the query, for example, `Roosevelt Avenue' is an avenue across Flushing, NY. In comparison, entities from irrelevant documents are much noisier. 
\texttt{Qw-De} features make use of this information and generate useful ranking evidence.

\begin{table}[t]
\centering
\caption{Query annotation accuracy and the gain of attention mechanism.
TagMe Accuracy includes the precision and recall of TagMe on ClueWeb queries, evaluated in prior research~\cite{Xiong2016BOE}. Attention Gains are the relative improvements of \texttt{AttR-Duet} compared with \texttt{LeToR-All}. Statistical significant gains are marked by $\dagger$.
\label{tab:att_gain}
}
\begin{tabular}{l|cc||cc}
\hline
& \multicolumn{2}{c||}{\bf{TagMe Accuracy}} &
\multicolumn{2}{c}{\bf{Attention Gain}} \\ \hline
& \textbf{Precision} & \textbf{Recall} & \textbf{NDCG@20}  & \textbf{ERR@20}  \\ \hline
\textbf{ClueWeb09} & 0.581 & 0.597 &  $ {+3.16\%}^{\dagger} $ & $ +3.65\%  $ \\ \hline
\textbf{ClueWeb12} & 0.460 & 0.555 &  $ {+14.20\%}^{\dagger}  $ &  $ {+15.45\%}^{\dagger} $ \\
\hline
\end{tabular}
\end{table}

\begin{figure}[t]
\centering
\begin{subfigure}{0.49\linewidth}
\includegraphics[width=1\linewidth]{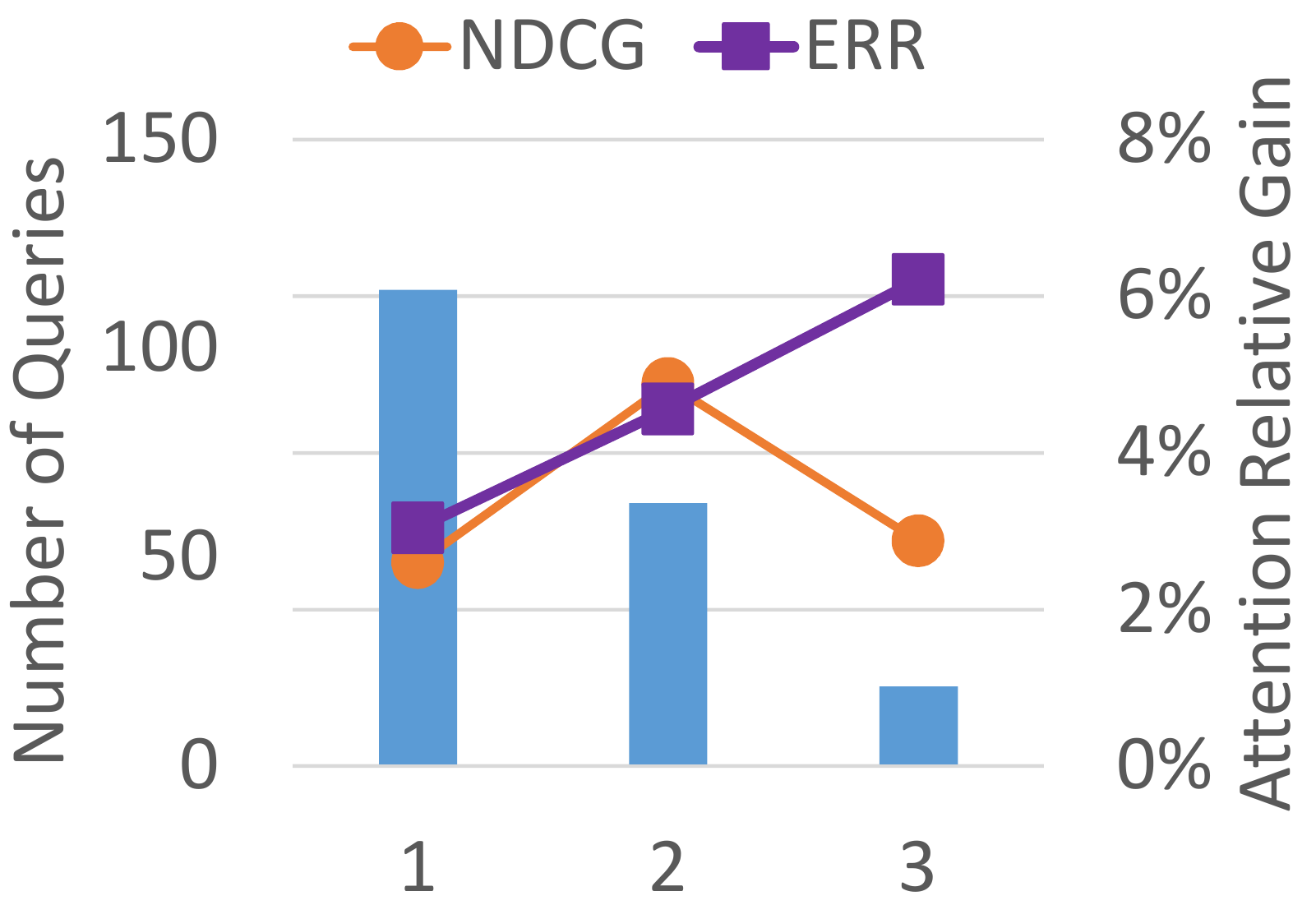}
\caption{ClueWeb09-B
}
\end{subfigure}
\begin{subfigure}{0.49\linewidth}
\includegraphics[width=1\linewidth]{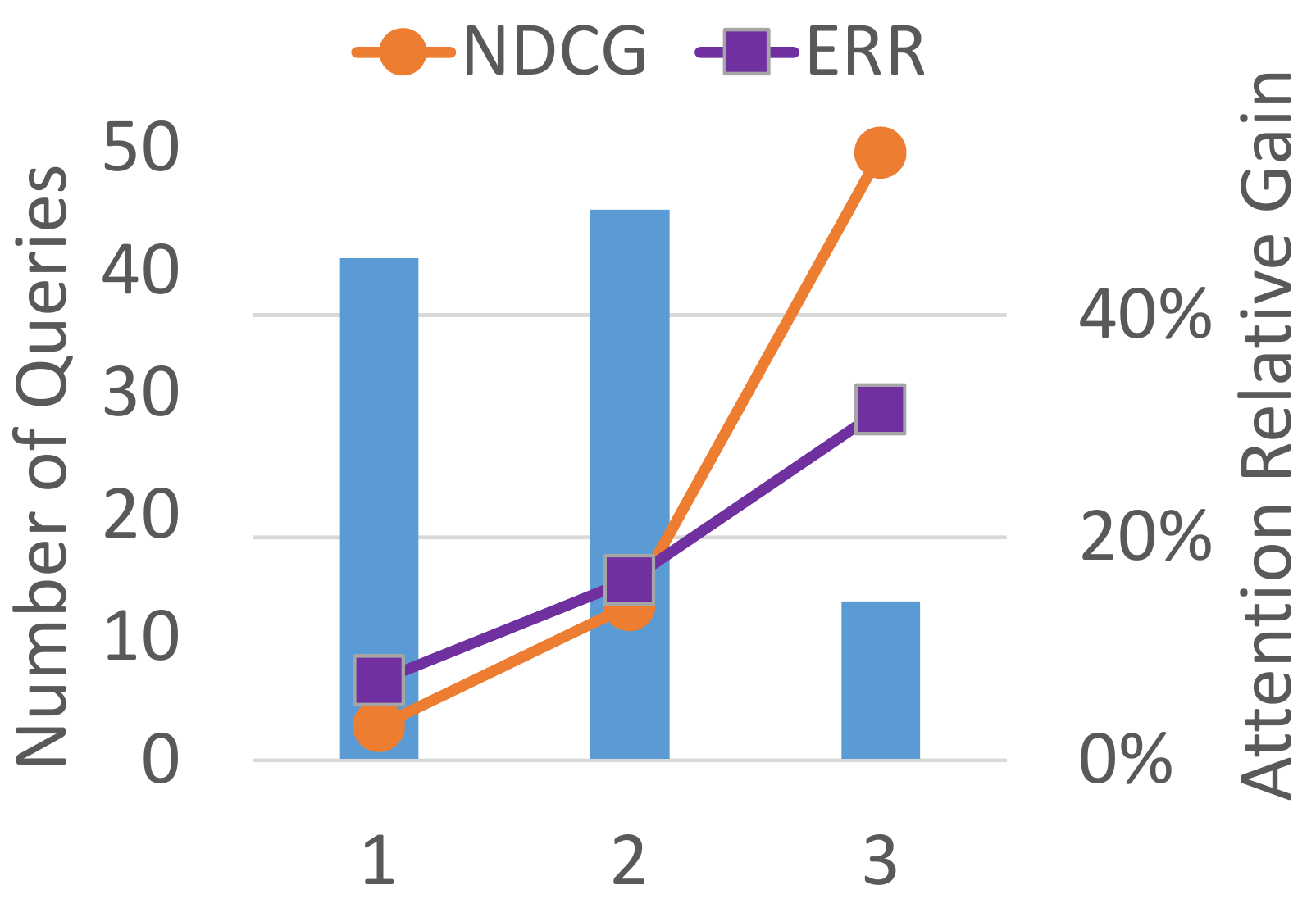}
\caption{ClueWeb12-B13
}
\end{subfigure}
\caption{Attention mechanism's gain on queries that contain different number of entities.
The x-axis is the number of entities in the queries. The y-axis is the number of queries in each group (histogram), and the gain from attention (plots). 
\label{fig:att_q_len}
}
\end{figure}

\textbf{\texttt{Qe-De} features} are extracted using the Explicit Semantic Ranking (ESR) method~\cite{ESR}. ESR is built upon the translation model. It operates in the entity space, and extracts the ranking features using histogram pooling. ESR was originally applied on scholar search. This work introduces ESR into web search and uses TransE model to train general domain embeddings from the knowledge graph.

To study the effectiveness of ESR in our setting, we also performed an incremental feature analysis of \texttt{LeToR-Qe-De}. Starting from the first bin (exact match), the following bins (soft matches) were incrementally added into RankSVM, and their rankings were evaluated.
The results are shown in Figure~\ref{fig:esr_eva}. The y-axis is the relative NDCG@20 over the base retrieval model they re-rank. The x-axis is the features used. For example, First 3 Bins refers to using the first three bins: $[1, 1]$, $[0.8, 1)$, and $[0.6, 0.8)$.

The observation on scholar search~\cite{ESR} holds on ClueWeb: Both exact match and soft match with entities are useful.
The exact match bin provided a 7\% improvement on ClueWeb09-B, while only 2\% on ClueWeb12-B13. 
 Similar exact match results were also observed in a prior study~\cite{Xiong2016BOE}. It is another reflection of the entity annotation quality differences on the two datasets.
Adding the later bins almost always improves the ranking accuracy, especially on ClueWeb12.

The second half of Table~\ref{tab:eg_feature} shows some examples of entities in the exact match bin and the nearest soft match bins. The exact match bin includes the query entities and is expected to help. The first soft match bin usually contains related entities. For example, the neighbors of `USS Yorktown (CV-10)' include `World War II' which is when the ship was built. The further bins are mostly background noise because they are too far away.  The improvements are mostly from the first 3 bins.

\subsection{Attention Mechanism Analysis~\label{sec:att_eva}}

\begin{table}[t]
\centering
\caption{Examples of learned attention. The entities in {\color{blue} \bf{bold blue}} draw more attention; those in {\color{mygray} gray} draw less attention.
\label{tab:eg_att}
}
\begin{tabular}{l|l}
\hline
\bf{Query} & \bf{Entity Attention} \\ \hline
\multirow{2}{*}{Balding Cure} & {\color{blue}\bf{`Cure'}} \\ 
 & {\color{mygray}`Clare Balding'} \\ \hline
\multirow{2}{*}{Nicolas Cage Movies} & {\color{blue} \bf{`Nicolas Cage'}}  \\
&{\color{mygray}`Pokemon (Anime)'} \\ \hline
Hawaiian Volcano  &  {\color{blue} \bf{`Volcano'}}; {\color{blue} \bf{`Observatory'}} \\
Observatories & {\color{mygray}`Hawaiian Narrative'}; \\ \hline
\multirow{2}{*}{Magnesium Rich Foods} & {\color{blue}\bf{`Magnesium'}};  {\color{blue} \bf{`Food'}} \\
&  {\color{mygray} {`First World'}} \\ \hline
{Kids Earth Day } &  {\color{blue} \bf{`Earth Day'}}  \\
Activities &  {\color{mygray} {`Youth Organizations in the USA'}} \\ \hline
\end{tabular}
\end{table}

The last experiment studies the effect of the attention mechanism by comparing \texttt{AttR-Duet} with \texttt{LeToR-All}. If enforcing flat attention weights on all query words and entities, \texttt{AttR-Duet} is equivalent to \texttt{LeToR-All}: The matching features, model function, and loss function are all the same. The attention part is their only difference, whose effect is reflected in this comparison.

The gains from the attention mechanism are shown in Table~\ref{tab:att_gain}. To better understand the attention mechanism's effectiveness in demoting noisy query entities, the query annotation's quality evaluated in a prior work~\cite{Xiong2016BOE} is also listed. The percentages in the Attention Gain columns are relative improvements of \texttt{AttR-Duet} compared with \texttt{LeToR-All}. $\dagger$ marks statistical significance. Figure~\ref{fig:att_q_len} breaks down the relative gains to queries with different numbers of query entities. The x-axis is the number of query entities. The histograms are the number of queries in each group, marked by the left y-axis. The plots are the relative gains, marked by the right y-axis.

The attention mechanism is essential to ClueWeb12-B13. Without the attention model, \texttt{LeToR-All} was confused by the noisy query entities and could not provide significant improvements over word-based models, as discussed in the last experiment.
With the attention mechanism, \texttt{AttR-Duet} improved \texttt{LeToR-All} by about $15\%$, outperforming all baselines. 
On ClueWeb09 where TagMe's accuracy is better~\cite{Xiong2016BOE}, the ranking evidence from the word-entity duet was clean enough for \texttt{LeToR-All} to improve ranking, so the attention mechanism's effect was smaller. Also, in general, the attention mechanism is more effective when there are more query entities,  while if there is only one entity there is not much to tweak.

The motivation for using attention is to handle the uncertainties in the query entities, a crucial challenge in utilizing knowledge bases in search. These results demonstrated its ability to do so. We also found many intuitive examples in the learned attention weights, some listed in Table~\ref{tab:eg_att}. The {\color{blue} \bf{bold  blue}} entities on the first line of each block gain more attention ($> 0.6$ attention score). Those in {\color{mygray} gray} on the second line draw less attention ($<0.4$  score). The attention mechanism steers the model away from those mistakenly linked query entities, which makes it possible to utilize the correct entities' ranking evidence from a noisy representation.

\newpage

%% file: Conclusion.tex
\section{Conclusions and Future Work}
This work presents a word-entity duet framework for utilizing knowledge bases in document ranking. In this paper, the query and documents are represented by both word-based and entity-based representations. The four-way interactions between the two representation spaces form a word-entity duet that can systematically incorporate various semantics from the knowledge graph. From query words to document words (\texttt{Qw-Dw}), word-based ranking features are included. From query entities to document entities (\texttt{Qe-De}), entity-based exact match and soft match evidence from the knowledge graph structure are included. The entities' textual fields are used in the cross-space interactions \texttt{Qe-Dw}, which expands the query, and \texttt{Qw-De}, which enriches the document.

To handle the uncertainty introduced from the automatic-thus-noisy entity representations, a new ranking model \texttt{AttR-Duet} is developed.
It employs a simple attention mechanism to demote the ambiguous or off-topic query entities, and learns simultaneously how to weight entities of varying quality and how to rank documents with the word-entity duet.

Experimental results on the TREC Web Track ad-hoc task demonstrate the effectiveness of proposed methods. \texttt{AttR-Duet} significantly outperformed all word-based and entity-based ranking baselines on both ClueWeb corpora and all evaluation metrics. Further experiments reveal that the strength of the method comes from both the advanced matching evidence from the word-entity duet, and the attention mechanism that successfully `purifies' them. On ClueWeb09 where the query entities are cleaner, all the entity related matching components from the duet provide similar or better improvements compared with word-based features. On ClueWeb12 where the query entities are noisier, the attention mechanism steers the ranking model away from noisy entities and is necessary for stable improvements.

Our method provides a unified representation framework to utilize knowledge graphs in information retrieval. 
As the first step, 
this work kept its components as simple as possible. It is easy to imagine further developments in various places. For example, the recent approaches in neural ranking with word embeddings can be incorporated~\cite{jiafeng2016deep}; better knowledge graph embeddings can be used~\cite{TransR}; better entity search methods can be applied when extracting word to entity features~\cite{Jing2016LTR}; the attention mechanism can be extended to document's entity-based representations. 
More sophisticated neural ranking models~\cite{K-NRM} can also be applied with the word-entity duet, especially when more training data are available.  

\section{Acknowledgments}
This research was supported by National Science Foundation (NSF) grant IIS-1422676, a Google Faculty Research Award, and a fellowship from the Allen Institute for Artificial Intelligence. We thank Xu Han for helping us train the TransE embedding.
Any opinions, findings, and conclusions in this paper are the authors' and do not necessarily reflect those of the sponsors. 
% \newpage